\documentclass[12pt]{article}
\usepackage{amsfonts,tcom}
\eqsecnum

\title{On relativistic models in the equilibrium statistical mechanics}
\author{V.~Tretyak}
\date{\small Institute for Condensed Matter Physics\\
of the Ukrainian National Academy of Sciences,\\
1~Svientsitskyy St., Lviv, Ukraine UA--290011\\
E-mail: tretyak{\tt@}icmp.lviv.ua}

\begin{document}
\maketitle
\begin{abstract}
Relativistic effects in the thermodynamical properties
of interacting particle systems are investigated within the framework
of the relativistic direct interaction theory in various forms of
dynamics. In the front form of relativistic dynamics an exactly solvable
model of a one-dimensional hard spheres gas is formulated and an
equation of state and thermodynamical potentials for such a gas are found.
Weakly-relativistic corrections to the thermodynamical functions of the
dilute gas with short-range interactions are discussed on the basis of the
approximately relativistic Hamiltonian function in the instant form of
dynamics.
\end{abstract}

\section{Introduction}

The present status of the relativistic direct interaction theory
\cite{[1],[2],[3],[4]}
enables us to consider it as a natural basis for a consistent
description of relativistic effects in various physical systems on
different levels: classical, quantum, or statistical ones.
The relativistic statistical mechanics of interacting particle systems
is nowadays at an early stage of its development, although the classical
partition function of a relativistic ideal gas was calculated by J\"uttner
in 1911. One cause of this was indicated in report \cite{thw}. Ter Haar
and Wergeland wrote: ``At extremely high temperatures relativistic effects
may, of course, be important. Then, however, matter behaves as a mixture of
ideal gases and this limiting case poses no problem. By and large, a
relativistic theory of heat seems, therefore, to be of little practical
importance''.  But the `practical importance', being a very nonsmooth
function on historical time, cannot be considered as the main reason
for theoretical investigation.

Among various approaches to the relativistic direct interaction theory the
single-time Lagrangian formalism \cite{[1],[5],[8]}
seems to be the most convenient in the consideration of the general problem
of relativistic dynamics, as well as in the investigation of various
approximations. This formalism has been extended to an arbitrary form of
relativistic dynamics defined geometrically by means of space-like foliations
of the Minkowski space \cite{[9],[7]}. Transition from the classical
Lagrangian to the Hamiltonian description allows one to consider the
relativistic effects in the statistical and quantum mechanical properties
of the particle systems.

The present paper is concerned with relativistic models of the
equilibrium statistical mechanics. In Sec.~2, a brief introduction
into the concept of the form of relativistic dynamics is presented.
Investigation of a classical and quantum relativistic ideal gas by
means of the front form of dynamics is outlined in Sec.~3.
Relativistic generalization of a one-dimensional gas of hard spheres is
obtained in Sec.~4. This gives an example of an exactly solvable model
in the relativistic statistical mechanics with a nontrivial particle
interaction. Sec.~5 is devoted to the investigation of the first
quasirelativistic (post-Newtonian) approximation. In the
instant form of dynamics the weakly-relativistic corrections to the
thermodynamical functions of the dilute gas with short-range interactions
are studied on the basis of the general structure, to the order $c^{-2}$, of
the approximately relativistic Hamiltonian. As an example, the relativistic
correction to the Van der Waals equation to the order $c^{-2}$ is obtained.

\section{Forms of relativistic dynamics in the La\-gran\-gian
description of an interacting particle system}

Let us consider a dynamical system consisting of $N$ interacting point
particles. It is convenient to describe the evolution of this system in
the ($n+1$)-dimensional Minkowski space $\mathbb M_{n+1}$
with coordinates $x^{\mu}$, $\mu=0,1,\ldots,n$.
We use the metric
$\|\eta _{\mu \nu}\|={\rm diag}(1,\underbrace {-1,\cdots ,-1}_{n})$.
In applications we put $n=3$ or $n=1$. The motion of the
particles is described by the world lines
\begin{equation}\label{2.1}
\gamma _a:\mathbb R\rightarrow\mathbb M_{n+1},\quad
\tau _{a}\mapsto x_{a}^{\mu}(\tau _{a});\qquad
a=1,\ldots ,N,
\end{equation}
being time-like one-dimensional unbounded submanifolds of the
Minkowski space.

The relativistic freedom in the simultaneity definition makes possible
different three-dimensional descriptions of the relativistic particle
motions. According to Dirac \cite{[10],[13]}, they are called the
{\it forms of relativistic dynamics\/}.
Within the framework of the single-time Lagrangian or Hamiltonian
mechanics this concept may be introduced in the following
way \cite{[9],[8]}. Let us consider foliation $\Sigma $ of the Minkowski
space $\mathbb M_{n+1}$ by the hypersurfaces
\begin{equation}\label{2.2}
t=\sigma (x),\qquad t\in \mathbb R,
\end{equation}
with the next property: every hypersurface
$\Sigma _t = \{ x\in \mathbb M_{n+1}~|~\sigma (x)=t\}$
must intersect the world lines $\gamma _a$ of all the particles
in one and only one point
\begin{equation}\label{2.3}
x_a(t)=\gamma _a\bigcap \Sigma_t.
\end{equation}
This allows us to consider $t$ as an evolution parameter of the system
\cite{[9],[13]}. In the Poincar\' e-invariant theory, when we consider only
time-like world lines, the hypersurfaces (\ref{2.2}) must be space-like or
isotropic:
\begin{equation}\label{2.4}
\eta _{\mu \nu}(\partial ^{\mu}\sigma )(\partial ^{\nu}\sigma )\geq 0,
\end{equation}
where $\partial ^{\mu}=\partial /\partial x_{\mu}$.
Then, we have $\partial ^0\sigma>0$, and the hypersurface equation
\re{2.2} has the solution $x^0=\varphi (t,{\bf x})$,
where ${\bf x}=(x^i)$, $i=1,\ldots ,n$.
Therefore, the constraint $x_a(t)\in \Sigma _t$ enables us to determine
the zeroth component of $x_a(t)$ in terms of $t$ and $x_a^i(t)$.
The parametric equations \re{2.1} of the world lines of
the particles in the given form of dynamics have the form:
\begin{equation}\label{2.5}
x^0=\varphi (t,{\bf x}_a(t))\equiv \varphi _a,\qquad
x^i=x^i_a(t).
\end{equation}

The evolution of the system is determined by  $nN$ functions
$t\mapsto x_a^i(t)$. They may be considered as representatives for
the sections $s:\mathbb R\rightarrow\mathbb F,~t\mapsto (t,x_a^i(t))$ of
the trivial fibre bundle $\pi:\mathbb F\rightarrow \mathbb R$ with
$nN$-dimensional fibre space $\mathbb E=\mathbb R^{nN}$.
The latter constitutes the configuration space of our system.

Three Dirac's forms of relativistic dynamics correspond to the following
hypersurfaces \re{2.2}: $x^0=ct$ (instant form), $x^0+x^n=ct$ (front form),
and $\eta _{\mu \nu}x^{\mu}x^{\nu} =c^2t^2$ (point form). Other
examples for the case $n=3$ may be found in \cite{[9]}.

In the relativistic Lagrangian mechanics
the Lagrangian function $L:J^{\infty}\pi \rightarrow \mathbb R$ is defined
on the infinite order jet space of the fibre bundle
$\pi :\mathbb F\rightarrow \mathbb R$ with the standard coordinates
$x_a^{i(s)}$ \cite{[14]}. The values of these coordinates for the section
$s:t\mapsto (t,x_a^i(t))$ belonging to the corresponding equivalence
class from $J^{\infty}\pi$ are
$x_a^{i(s)}(t)=d^sx_a^i(t)/dt^s\equiv D^sx_a^i$, $s=0,1,2,\ldots$.

The free-particle system is determined in any form of dynamics by the
Lagrangian $L_f:J^1\pi \to \mathbb R$ depending on the first derivatives:
\begin{equation}\label{2.7}
L_f=\sum _a m_ac \sqrt{\left (D\varphi (t,{\bf x})\right )^2-v_a^2 }; \qquad
v_a^i\equiv x_a^{i(1)},
\end{equation}
$m_a$ being a rest mass of the particle.
In the front form of dynamics it reads:
\begin{equation}\label{2.6b}
L_f=\sum_a m_ac^2 \sqrt{1-(v_{a1}^2+\cdots +v_{a(n-1)}^2)/c^2-
2v_{an}/c}.
\end{equation}

In the general case the Poincar\' e-invariance conditions forbid the
existence  of interaction Lagrangians which are defined on the jet-space
$J^{r}\pi $ with some finite $r$ (for example, with $r=1$). This leads to
serious difficulties in physical interpretation of the formalism, and,
in fact, makes it impossible to obtain a closed form of the corresponding
Hamiltonian functions which is necessary for the development of the classical
statistical mechanics \cite{[1],[8],[9]}. This fact is the Lagrangian
counterpart of the famous no-interaction theorem in the Hamiltonian
relativistic mechanics \cite{[15]}.

There are at least two possibilities to avoid this
difficulty. The first is offered by the front form of dynamics in the
two-dimensional Minkowski space. In this case there exists a wide class
of interaction Lagrangians depending on the first order derivatives
\cite{[11]}. This allows us to obtain corresponding Hamiltonian
functions by the standard Legendre transformation that preserves the physical
meaning of the position canonical coordinates. The second consists in the
consideration of the approximation in $c^{-2}$ \cite{[1],[7],[12]}.
At least in the
first order approximation we may obtain the usual Lagrangian functions
depending on the first order derivatives \cite{[1]}.
Examples of using these possibilities in the investigation of
relativistic effects in models of the equilibrium statistical mechanics
will be considered in the next sections.

\section{Relativistic ideal gas within the framework of the front form of
dynamics}

As an illustrative example we shall consider here a description of the
relativistic ideal gas by means of the front form of dynamics.
The Hamiltonian function of the free $N$-particle system has the form:
\begin{equation}\label{3.1}
H=\sum _{a=1}^NH_a({\bf x}_a,{\bf p}_a)
\end{equation}
with
\begin{equation}\label{3.2}
H_a=H_0({\bf p})=c\frac{m^2c^2+{\bf p}^2_a}{2p_{a3}}.
\end{equation}
This Hamiltonian can be obtained from the Lagrangian \re{2.6b} by the
usual Legendre application.

The $n$-dimensional coordinates ${\bf x}_a$ cover the given region
$\Omega \subset \mathbb R^n$
with ${\rm vol}(\Omega )=V$, and ${\bf p}_a$ belongs to the region
${\mit\Pi}=\{ {\bf p}_a=(p_{a1},\ldots,p_{an})\in \mathbb R^n~|~p_{an}> 0 
\}$.As it is well known, all the information about thermodynamical properties
of the system is contained in the expression for the canonical partition
function \cite{[16]}:
\begin{equation}\label{3.3}
Z_N=\frac {1}{h^{nN}N!}\int e^{-\beta H}\prod _{a}d^nx_ad^np_a,
\end{equation}
where, as usual, $\beta = (kT)^{-1}$ and the integration is performed over
the phase space $\mathbb P$. In our case $\mathbb P=\Omega^N\times {\mit 
\Pi}^N$.Inserting \re{3.1} into Eq.~\re{3.3} we obtain
\begin{equation}\label{3.4}
Z_N=\frac {1}{h^{nN}N!}z^N,
\end{equation}
where
\begin{equation}\label{3.5}
z=\int e^{-\beta H_a}d^nx_ad^np_a.
\end{equation}
Substitution of the front form Hamiltonian \re{3.2} immediately yields
\begin{equation}\label{3.6}
z=V\int \limits_0^{\infty} dp_ne^{-\frac 12\beta c \left (
p_n+\frac {m^2c^2}{p_n}\right )}
\prod_{k=1}^{n-1} \int\limits_{-\infty }^{\infty }dp_k
e^{-\frac {\beta c}{2p_n}p_k^2}.
\end{equation}
Performing integration over $p_k$ and putting $p_n=mc\alpha $ we have
\begin{equation}\label{3.7}
z=Vmc\left (\frac {2\pi m}{\beta}\right )^{(n-1)/2}
\int \limits_0^{\infty} d\alpha \alpha ^{(n-1)/2}e^{-\frac {\beta mc^2}{2}
\left (\alpha +\frac {1}{\alpha }\right )}.
\end{equation}
Using the integral representation of the Hankel function
$K_n(x)=\frac {\pi i}{2}e^{in\pi /2}H_n^{(1)}(ix)$,
\begin{equation}\label{3.8}
K_n(x)=\frac12\int \limits_0^{\infty} d\alpha \alpha ^{n-1}e^{-\frac x2
\left (\alpha +\frac {1}{\alpha }\right )}, \qquad x>0,
\end{equation}
we obtain
\begin{equation}\label{3.9}
z=2Vmc\left (\frac {2\pi m}{\beta}\right )^{(n-1)/2}K_{(n+1)/2}
(\beta mc^2).
\end{equation}
When $n=3$, Eqs.~(\ref{3.4}) and (\ref {3.9}) give well-known J\"uttner's
result:
\begin{equation}\label{3.10}
Z_N=\frac {1}{N!}\left [\frac {4\pi m^2cV}{\beta h^3}K_2(\beta mc^2)\right ]^N,
\end{equation}
which is usually derived by means of the instant form Hamiltonian:
\begin{equation}\label{3.11}
H_a=c\sqrt {m^2c^2+{\bf p}_a^2}.
\end{equation}

Using the asymptotical expansion
\begin{equation}\label{3.12}
K_n(x)\asymp \sqrt{\frac {\pi}{2x}}e^{-x}\left (1+\frac{4n^2-1}{8x}+\ldots
\right ),\qquad x\to \infty ,
\end{equation}
we can obtain a weakly-relativistic correction to the nonrelativistic
result:
\begin{equation}\label{3.13}
Z_N=Z_N^{(0)}e^{-\beta Nmc^2}\left (1+\frac{n(n+2)}{8\beta mc^2}\right )^N,
\end{equation}
where
\begin{equation}\label{3.14}
Z_N^{(0)id}=\frac{V\Lambda ^{-nN}}{N!}, \qquad
\Lambda \equiv \sqrt{\frac {\beta h^2}{2\pi m}}.
\end{equation}
The necessity for (rather trivial) renormalization of the nonrelativistic
partition function $Z_N^{(0)}\to Z_N^{(0)}e^{-\beta Nmc^2}$ follows from
the presence of the rest energy $mc^2$ in the relativistic Hamiltonians
\re{3.2} or \re{3.11}.

The obtained expressions can be useful for the treatment of the
quantum relativistic ideal gas with generalized statistics of arbitrary
order $q$. In this case it is convenient to consider
the grand partition function
\begin{equation}\label{gr}
\Theta ={\rm Tr}e^{-\beta ({\sf H}-\mu{\sf N})}\equiv e^{-\beta \Omega},
\end{equation}
where ${\sf H}$ and ${\sf N}$ represent the operators of energy and
of particle number, respectively, and $\mu$ denotes chemical potential
\cite{[16]}.

Assuming that the maximum value of the occupation number in a given
energy state can be $q$, any integer greater than 1, and
acting in a manner quite similar to the nonrelativistic case (cf.,
for example, \cite{indian}), we obtain:
\begin{equation}\label{3.17}
\Omega =-\frac {gV}{\beta h^{n}}\int d^np\ln\left(\frac{1-
e^{-\beta (q+1)(H_0({\bf p})-\mu )}}{1-e^{-\beta (H_0({\bf p})-\mu )}}
\right).
\end{equation}
Here $g$ denotes degeneracy of the energy state with a given value of
momentum ${\bf p}$. For structureless particles with the spin
$\sigma$ we have $g=2\sigma +1$.

Of course, care is needed in the replacement of the summation by the
integration over the momentum states, when $q\sim N$, and, especially,
if $q\to \infty $ (that corresponds to the Bose--Einstein statistics). But
this problem, connected with the Bose condensation, is essential at a
low temperature, $\beta ^{-1}<<mc^2$, when relativistic correction
seems to be negligible. Then the standard thermodynamical relations
lead to the following expression for the pressure:
\begin{equation}\label{3.18}
P=g\beta^{-1}h^{-n}\int d^np\left[\ln\left(1-\lambda^{q+1}
e^{-\beta (q+1)H_0({\bf p})}\right)
-\ln \left(1-\lambda e^{-\beta H_0({\bf p})}\right)
\right],
\end{equation}
where
\begin{equation}\label{3.16}
\lambda =e^{\beta \mu}
\end{equation}
is the fugacity. The number density
\begin{equation}\label{3.19}
\frac 1v=\frac NV=\frac{\partial P}{\partial \mu}
\end{equation}
is given by
\begin{equation}\label{3.20}
\frac 1v=gh^{-n}\int d^np\left[\frac{\lambda e^{-\beta H_0({\bf p})}}
{1-\lambda e^{-\beta H_0({\bf p})}}-(q+1)
\frac{\lambda^{q+1}e^{-\beta (q+1)H_0({\bf p})}}{1-
\lambda^{q+1} e^{-\beta (q+1)H_0({\bf p})}}
\right]
\end{equation}
and the inner energy density $u=U/V$ is found to be
\begin{equation}\label{3.21}
u=gh^{-n}\int d^npH_0({\bf p})\left[\frac{\lambda e^{-\beta H_0({\bf p})}}
{1-\lambda e^{-\beta H_0({\bf p})}}-
(q+1)\frac{\lambda^{q+1}e^{-\beta (q+1)H_0({\bf p})}}{1-
\lambda^{q+1} e^{-\beta (q+1)H_0({\bf p})}}\right].
\end{equation}
Expanding the exponential functions into the series and using the
denotation
\begin{equation}\label{3.22}
\phi (\beta )=gh^{-n}z/V,
\end{equation}
where $z$ is defined by Eq.~(\ref{3.5}), we have
\begin{equation}\label{3.23}
\beta P=\sum_{k=1}^{\infty}k^{-1}\left[\lambda ^k\phi(\beta k)-
\lambda ^{k(q+1)}\phi(\beta (q+1)k)
\right]
\end{equation}
and
\begin{equation}\label{3.24}
\frac 1v=\sum_{k=1}^{\infty}\left[\lambda ^k\phi(\beta k)-(q+1)
\lambda ^{k(q+1)}\phi(\beta (q+1)k)
\right].
\end{equation}
Keeping only the terms up to the second order in the density $1/v$, we
obtain an equation of state of the form:
\begin{equation}\label{3.25}
\beta P=\frac 1v-\frac 1{2v^2}\eta_q\frac{\phi (2\beta)}{\phi (\beta)},
\end{equation}
where
\begin{eqnarray}\label{3.26}
\eta_q=\left \{ \begin {array}{cc}-1, & q=1 \\
1 , & q>1. \end {array} \right.
\end{eqnarray}
Similarly, for the energy density with the same accuracy we have
\begin{equation}\label{3.27}
u=-\frac 1v\frac{\phi '(\beta)}{\phi (\beta)}+
\frac 1{2v^2}\eta_q\frac{\phi (2\beta)\phi '(\beta)-
\phi (\beta)\phi'(2\beta)}{\phi (\beta)^3}.
\end{equation}
Making use of the explicit expression for the $\phi $ function from
\re{3.9} and \re{3.22}, we may find a relativistic equation of state at a
high temperature up to the first order in the degeneracy parameter
\begin{equation}\label{3.28}
\delta=\frac{\Lambda^n}{gv}.
\end{equation}
The equation is found to be
\begin{equation}\label{3.29}
\beta P=\frac 1v\left[1-\frac{\eta_q\delta}{2^{1+n/2}}F(\beta mc^2)
\right],
\end{equation}
where
\begin{equation}\label{3.30}
F(x)=\sqrt{\frac{\pi}x}\frac{K_{(n+1)/2}(2x)}{K^2_{(n+1)/2}(x)}.
\end{equation}
Because $F(x)\to 1$ as $x\to \infty$, Eq.~\re{3.29} agrees with the known
nonrelativistic results \cite{[16]}. In a similar manner we may write
relativistic virial expansions for other thermodynamical functions of the
quantum ideal gas, especially for the inner energy and specific
heat.

\section{Relativistic one-dimensional model of the hard spheres gas}

In the two-dimensional space-time $\mathbb M_2$, the front form
of dynamics serves a wide class of nontrivial interaction
Hamiltonians in terms of the covariant canonical coordinates.

The general form of a Hamiltonian function for a system of $N$
identical particles on the line $\mathbb I=\{x\in \mathbb R~|~0<x<A\}$
described by the canonical coordinates $x_a$ and momenta $p_a$ within the
framework of the front form of dynamics is given by \cite{[10]}:
\begin{equation}\label{4.1}
H=\sum _aH_0(p_a)+\dsum ab (p_a+p_b)V(r_{ab}p_a,r_{ab}p_b).
\end{equation}
Here
\begin{equation}\label{h0}
H_{0}(p)=\frac 12\left (p+\frac {m^2}{p}\right ) ,
\end{equation}
$r_{ab}\equiv x_a-x_b$, and V is an
arbitrary function on the indicated arguments. In this section we put $c=1$.
From the definition of the front form of dynamics it follows that momenta
$p_a$ belong to the positive semiaxis $\mathbb R_+=\{x\in \mathbb R~|~x>0\}$:
\begin{equation}\label{4.3}
p_a>0.
\end{equation}
For the convenience of the comparison with the corresponding nonrelativistic
calculations we assume that function $V$ has the form:
\begin{equation}\label{4.4}
V=V(|q_{ab}|) ,
\end{equation}
where
\begin{equation}\label{4.5}
q_{ab}=r_{ab}(p_a+p_b)\nu \left (\frac {p_a}{p_b}+\frac {p_b}{p_a}\right ) ,
\end{equation}
and $\nu :\mathbb R_+\rightarrow \mathbb R_+$ is some function which
will be defined later.

Let us again consider the canonical partition function
\begin{equation}\label{4.6}
Z_N=\frac {1}{h^NN!}\int e^{-\beta H}\prod _{a}dx_adp_a ,
\end{equation}
over the phase space $\mathbb P=\mathbb I^N\times \mathbb R_+^N$.

Since we are interested in the relativistic generalization of the
hard spheres model, we can choose interaction function \re{4.4}
in the form
\begin{eqnarray}\label{4.7}
V(x)=\left \{ \begin {array}{cc}0, & x>\sigma \\
\infty , & x<\sigma . \end {array} \right.
\end{eqnarray}
We, therefore, find
\begin{eqnarray}\label{4.8}
e^{-\beta V(|q_{ab}|)}=\left \{ \begin {array}{cc}1, & |q_{ab}|>\sigma \\
0 , & |q_{ab}|<\sigma  .\end {array} \right.
\end{eqnarray}
If we choose function $\nu (x)$ in such a way that for any $a,b,c$
\begin{equation}\label{4.9}
{\rm if}\quad |q_{ab}|>\sigma \quad {\rm and} \quad|q_{bc}|>\sigma ,
\quad {\rm then}\quad |q_{ac}|>\sigma ,
\end{equation}
the partition function \re{4.6} can be written in the form:
\begin{equation}\label{4.10}
Z_N=\frac {1}{h^NN!}\int \prod _{a}d\mu (p_a)\int \prod _{a}dx_a ,
\end{equation}
where
\begin{equation}\label{4.11}
d\mu (p)=e^{-\beta H_0(p)}dp
\end{equation}
and coordinates $x_a$ must satisfy the restrictions
\begin{equation}\label{4.12}
|r_{ab}|>\frac {\sigma }{p_a+p_b}
u \left (\frac {p_a}{p_b}+\frac {p_b}{p_a}\right ) , \qquad
u(x)\equiv \frac {1}{\nu (x)} .
\end{equation}
Conditions \re{4.9} lead to the inequality
\begin{equation}\label{4.13}
\frac x4\leq \frac {u(x)}{u(2)}\leq \frac {x+2}{4} .
\end{equation}
In the following we shall choose the simplest solution of \re{4.13}
which has the form
\begin{equation}\label{4.14}
u(x)=C(x+2) ,
\end{equation}
such that
\begin{equation}\label{4.15}
q_{ab}=C^{-1}r_{ab}\left (\frac {1}{p_a}+\frac {1}{p_b}\right )^{-1} .
\end{equation}
Because in the nonrelativistic limit $p_a \rightarrow m$, the demand that
$q_{ab} \rightarrow r_{ab}$ in this limit fixes the value of the constant
$C$:
\begin{equation}\label{4.16}
C=\frac m2.
\end{equation}
Therefore, partition function (\ref{4.10}) may be rewritten in the form:
\begin{equation}\label{4.17}
Z_N=\frac {1}{h^N}\int \left (\prod _{a=1}^{N}d\mu (p_a)dx_a\right )
\prod _{a=1}^{N-1}\theta \left (x_{a+1}-x_a-
\frac {m\sigma}{2p_a}-\frac {m\sigma}{2p_{a+1}}\right ) .
\end{equation}
The appearance of Heaviside $\theta$ functions is a consequence of
conditions (\ref{4.12}). Let us perform in \re{4.17} a change of
the variables $(x_a,p_a)\rightarrow (y_a,p_a)$, such that the arguments of
$\theta$ functions become $y_{a+1}-y_a$. This gives
\begin{equation}\label{4.18}
y_a=x_a-\frac {m\sigma}{2}\varphi _a ,
\end{equation}
where
\begin{equation}\label{4.19}
\varphi _1=0,\qquad \varphi _2=\frac {1}{p_1}+\frac {1}{p_2} ,
\end{equation}
and
\begin{equation}\label{4.20}
\varphi _a=\frac {1}{p_1}+
2\sum _{b=2}^{a-1}\frac {1}{p_b}+\frac {1}{p_a}, \qquad a>2 .
\end{equation}
As a consequence, we obtain
\begin{equation}\label{4.21}
Z_N(A, \beta )=\frac {1}{h^NN!}\int \prod _{a=1}^{N}d\mu (p_a)
\theta \left (A-\frac {m\sigma}{2}\varphi _N\right )
\left (A-\frac {m\sigma}{2}\varphi _N\right )^N .
\end{equation}
Before performing an integration over momenta we consider the Laplace
transformation of \re{4.21}
\begin{eqnarray}\label{4.22}
Z_N(s,\beta)&=&\int \limits_{0}^{\infty}dAe^{-sA}Z_N(A,\beta)\nonumber\\
&=&\frac {1}{h^Ns^{N+1}}\int \prod _{a=1}^{N}d\mu (p_a)
\exp \left (-s\frac {m\sigma}{2}\varphi _N\right) .
\end{eqnarray}
Using expressions \re{4.11} and \re{4.19}, \re{4.20} we find
\begin{equation}\label{4.23}
Z_N(s,\beta)=\frac {1}{h^Ns^{N+1}}z(\beta ,M_2)^2z(\beta ,M_1)^{N-2} .
\end{equation}
where
\begin{equation}\label{4.24}
M_1^2=m^2+2s\frac {m\sigma}{\beta},\qquad
M_2^2=m^2+s\frac {m\sigma}{\beta} ,
\end{equation}
and
\begin{equation}\label{4.25}
z(\beta ,m)=\int \limits_{0}^{\infty}dp\exp \left [-\frac {1}{2}\beta
\left (p+\frac {m^2}{p}\right )\right ]=2mK_1(\beta m) .
\end{equation}

Next, we consider the grand partition function
\begin{equation}\label{4.26}
Z(\beta ,s \lambda )=\sum _N\lambda ^NZ_N(\beta ,s).
\end{equation}
The summation over $N$ is performed immediately giving
\begin{equation}\label{4.27}
Z(\beta ,s, \lambda )=\frac {z^2(\beta M_2)}{z^2(\beta M_1)}
\frac {h}{sh-\lambda z(\beta M_1)}.
\end{equation}
The asymptotic behaviour of function  $Z(\beta ,A)$ is determined by the
singularity points of function \re{4.27} which lie on the real axis
for variable $s$. In our case there exists only one such a point, $s'$ ,
that is the solution to the equation
\begin{equation}\label{4.28}
s'h=\lambda z(\beta M_1(s')).
\end{equation}
Specifically, we have for the pressure $P=\beta ^{-1}s'$ \cite{[17]}.
Using \re{4.24} and \re{3.16} we obtain an expression for the
chemical potential in terms of $\beta$ and $P$:
\begin{equation}\label{4.29}
\mu (\beta, P)=-\frac {1}{\beta}\ln\frac {z(\beta M)}{h\beta P} ,
\end{equation}
where
\begin{equation}\label{4.30}
M^2=m^2+2m\sigma P .
\end{equation}
The standard thermodynamical relation \cite{[16]}
\begin{equation}\label{4.31}
d\mu =-\tilde sdT+vdP ,
\end{equation}
where $\tilde s=S/N$ and $v=A/N$ ,
allows one to obtain the equation of state
\begin{equation}\label{es}
v=\frac {\partial \mu}{\partial P}=
\frac {1}{\beta P}+\frac {m\sigma}{M}\frac {K_0(\beta M)}{K_1(\beta M)} .
\end{equation}

It is convenient to represent this equation in the form:
\begin{equation}\label{4.33}
v=\frac {1}{\beta P}+\sigma \delta (\beta ,P),
\end{equation}
where
\begin{equation}\label{4.34}
\delta (\beta ,P)=\frac mM\frac {K_0(\beta M)}{K_1(\beta M)}.
\end{equation}
From the inequality
\begin{equation}\label{ie}
K_n(x)>K_m(x), \qquad  {\rm if} \quad   n>m,
\end{equation}
it follows that
\begin{equation}\label{4.36}
0< \delta (\beta ,P)<1.
\end{equation}

In the nonrelativistic approximation, when $\beta m\to \infty $, $M \to m$,
the asymptotical expression (\ref{3.11}) shows that
$\delta \to 1$, and we obtain a well-known Tonks equation \cite{[18]}
\begin{equation}\label{4.37}
v=\frac {1}{\beta P}+\sigma .
\end{equation}

In the ultrarelativistic limit, when $\beta m\to  0$, we have $\delta \to 0$
and arrive at the state equation of the ideal gas. In the case of small
$\sigma $ (that corresponds to the linear approximation in the interaction)
we can replace function $\delta (\beta ,P)$ by its value $\delta _0(\beta )$
at $\sigma \to  0$:
\begin{equation}\label{4.38}
\delta _0=\frac {K_0(\beta m)}{K_1(\beta m)}.
\end{equation}
This quantity gets an interesting interpretation after accounting that
the mean value of $p^n$ over the free-particle distribution
defined by the Hamiltonian \re{h0} is given by
\begin{equation}\label{4.39}
\bra p^k\ket_0=\int \limits_{0}^{\infty}dpp^ke^{-\beta H_0(p)}
\left /
\int \limits_{0}^{\infty}dpe^{-\beta H_0(p)}\right.
=m^k\frac {K_{n+1}(\beta m)}{K_1(\beta m)}.
\end{equation}
Then $\delta _0(\beta )=m\bra p^{-1}\ket_0$. Next, we can observe that in the
front form of dynamics we have for free particles $p=m\gamma $, where
$\gamma ^{-1}=\sqrt{1-2v}$  corresponds to the Lorentz radical
$\sqrt{1-v^2}$. Therefore, in the linear approximation the state equation
\re{4.33} can be considered as a result of taking into account the well-known
Lorentz spatial contraction $\sigma \to \sigma \bra\gamma ^{-1}\ket_0$
in the nonrelativistic Tonks equation \re{4.37}.

From the equation of state \re{es} we get immediately
\begin{equation}\label{st}
\frac {\partial v}{\partial P}=
-\frac {1}{\beta P^2}-\beta \left (\frac {m\sigma}{M}\right )^2
\left[1-\frac {K_0^2(\beta M)}{K_1^2(\beta M)}\right ].
\end{equation}
Then inequality \re{ie} shows that the condition of thermodynamical
stability
\begin{equation}\label{4.41}
\left (\frac {\partial v}{\partial P}\right )_T<0
\end{equation}
is valid for equation \re{es} identically. Thus, the system does
not have any phase transition.

In a similar manner we may obtain explicit expressions for other
thermodynamical functions. For example, for entropy we have
\begin{eqnarray}\label{ent}
\tilde s&=&-\left (\frac {\partial \mu}{\partial T}\right )_P
=k\beta ^2\frac {\partial \mu}{\partial \beta}\nonumber\\
&=&k\left \{\ln \frac {2MK_1(\beta M)}{h\beta P}+2+
\beta M\frac {K_0(\beta M)}{K_1(\beta M)}\right \};
\end{eqnarray}
the energy density is determined by
\begin{equation}\label{4.43}
\tilde u=\mu+T\tilde s-Pv=\frac {1}{\beta}+
\frac {m^2+m\sigma P}M
\frac {K_0(\beta M)}{K_1(\beta M)}.
\end{equation}
Specific heat $c_P$ can be directly obtained from \re{ent} giving
\begin{eqnarray}\label{4.44}
c_P&=&T\left (\frac {\partial \tilde s}{\partial T}\right )_P
=-\beta \left (\frac {\partial \tilde s}{\partial \beta}\right )\nonumber\\
&=&k\left \{2+(\beta M)^2\left [1-\frac {K_0(\beta M)^2}{K_1(\beta M)^2}-
\frac {K_0(\beta M)}{\beta MK_1(\beta M)^2}\right ]\right \}.
\end{eqnarray}

Therefore, we have an exactly solvable example of a nontrivial particle
interaction in the relativistic statistical mechanics.

\section{Weakly-relativistic corrections to the thermodynamics of
an interacting particle system}

Here we shall consider a system of $N$ point-like particles with
pairwise interactions at the first post-Newtonian approximation.
Let the nonrelativistic interaction potential has the form:
\begin{equation}\label{5.1}
U^{(0)}=\dsum ab u_{ab};\qquad u_{ab}=u(r_{ab}),
\qquad r_{ab}\equiv |{\bf r}_{ab}|\equiv |{\bf x_a}-{\bf x}_b|.
\end{equation}
In that case the general form of the first post-Newtonian Lagrangian
function is given by \cite{[1]}:
\begin{eqnarray}\label{5.2}
&&L=\sum _a\left( \frac{m_av_a^2}{2}+\frac{m_av_a^4}{8c^2}\right )-
U^{(0)}\nonumber\\
&&{}+\frac {1}{2c^2}\dsum ab\left({\bf v}_a\cdot{\bf v}_bu_{ab}-
({\bf r}_{ab}\cdot{\bf v}_a)({\bf r}_{ab}\cdot{\bf v}_b)
\frac{1}{r_{ab}}\frac{du_{ab}}{dr_{ab}}\right)+c^{-2}\mit \Phi
\end{eqnarray}
with an arbitrary Galilei-invariant function $\mit \Phi$.
We choose this function in the form which is 
determined in terms of the nonrelativistic potential $u(r)$:
\begin{equation}\label{5.3}
\mit \Phi =\frac 12\dsum ab\left(Av_{ab}^2u_{ab}+
B({\bf r}_{ab}\cdot{\bf v}_{ab})^2
\frac{1}{r_{ab}}\frac{du_{ab}}{dr_{ab}}\right),
\quad {\bf v}_{ab}\equiv {\bf v_a}-{\bf v}_b,
\end{equation}
with two arbitrary numerical coefficients, $A$ and $B$.
This structure of the post-Newtonian Lagrangian is sufficient to
cover a wide class of interactions including those following from various
field-theoretical considerations. For example, the values
\begin{equation}\label{5.4a}
A=\sigma^2-1,\qquad B=0
\end{equation}
correspond to the interactions mediated by linear relativistic fields of
spin $\sigma $. Particularly, $\sigma=1$ and $u(r)\sim r^{-1}$ give the
famous Darwin's Lagrangian for electromagnetic interactions.
It is remarkable that expression \re{5.2} under \re{5.3} was derived by
Breit \cite{breit} as far back as 1937 with the aid of simple
symmetry treatment.

For the system of identical particles the Hamiltonian which follows
from the above Lagrangian has the form:
\begin{equation}\label{5.4}
H=H^{(0)}+c^{-2}H^{(1)}+O(c^{-4}),
\end{equation}
where $H^{(0)}$ is a nonrelativistic Hamiltonian,
\begin{equation}\label{5.5}
H^{(0)}=\sum _a\frac{p_a^2}{2m}+U^{(0)},
\end{equation}
and
\begin{eqnarray}\label{5.6}
H^{(1)}=-\sum _a\frac{p_a^4}{8m^3}-\frac 1{2m^2}\dsum ab\left \{ \left[
(1-2A){\bf p}_a\cdot{\bf p}_b +A(p_a^2+p_b^2)\right ]u_{ab}
\right.\nonumber \\
\left.-\left[(1+2B)({\bf r}_{ab}\cdot{\bf p}_a)({\bf r}_{ab}\cdot{\bf p}_b)
-B[({\bf r}_{ab}\cdot{\bf p}_a)^2+({\bf r}_{ab}\cdot{\bf p}_b)^2]\right ]
\frac{1}{r_{ab}}\frac{du_{ab}}{dr_{ab}}\right\}.
\end{eqnarray}
The canonical variables $({\bf x}_a,{\bf p}_a)$ are connected with the
Lagrangian $({\bf x}_a,{\bf v}_a)$ by the standard Legendre transformation
\begin{equation}\label{5.7a}
{\bf p}_a=\frac{\partial L}{\partial {\bf v}_a},
\end{equation}
considered to the order $c^{-2}$.
Inserting Hamiltonian \re{5.4} into expression \re{3.3} for
the classical partition function and expanding it to the $c^{-2}$ terms,
we get
\begin{equation}\label{6.3}
Z_N=\frac 1{h^{3N}N!}\int e^{-\beta H^{(0)}}
\left(1-\beta c^{-2}H^{(1)}\right)\prod _{a}d^3x_ad^3p_a
\equiv Z_N^{(0)}+c^{-2}Z_N^{(1)},
\end{equation}
where $Z_N^{0}$ is a nonrelativistic partition function
\begin{equation}\label{5.8}
Z_N^{0}=Z_N^{(0)id}Q,
\end{equation}
$Q$ being a nonrelativistic configuration integral
\begin{equation}\label{5.9}
Q=V^{-N}\int e^{-\beta U^{(0)}}
\left(1-\beta c^{-2}H^{(1)}\right)\prod _{a}d^3x_a.
\end{equation}

The straightforward calculation gives first-order corrections to the
nonrelati\-vis\-tic partition function \re{5.8}:
\begin{equation}\label{5.10}
Z_N^{(1)}=Z_N^{(0)}\frac{R}{\beta m};\qquad
Z_N=Z_N^{(0)}\left(1+\frac{R}{\beta mc^2}\right),
\end{equation}
where $R$ is defined in terms of the configuration integral (\ref{5.9}),
\begin{equation}\label{5.11}
R=\frac{15}8N-\frac3Q\left(A\beta \frac {\partial Q}{\partial \beta}+
BV\frac {\partial Q}{\partial V}\right).
\end{equation}
In the absence of interaction, when $Q=1$, the obtained expression agrees
with Eq.~\re{3.13}.

All thermodynamical properties of the system may be deduced from the
free energy:
\begin{equation}\label{5.12}
F=-\beta ^{-1}\ln Z=F^{(0)}-\frac R{\beta^2mc^2},
\end{equation}
where the corresponding nonrelativistic expression is given by
\begin{equation}\label{5.13}
F^{(0)}=-\beta ^{-1}\ln \left[\left(V\Lambda ^{-3}\right)^N\frac Q{N!}\right].
\end{equation}
Eq.~\re{5.12} can be rewritten in the form, which gives a first-order
correction in terms of the nonrelativistic free energy:
\begin{equation}\label{5.14}
F=F^{(0)}-\frac3{\beta mc^2}\left[\left(\frac58-\frac{3A}2+B\right)
\frac N{\beta}+A\left(F^{(0)}+\beta \frac {\partial F^{(0)}}{\partial \beta}
\right)+BV\frac {\partial F^{(0)}}{\partial V}\right].
\end{equation}
This formula may be also useful in the obtaining of weakly-relativistic
corrections to various phenomenological nonrelativistic results.

Let us consider corrections to the equation of state. In the
nonrelativistic limit it has the form:
\begin{equation}\label{5.15}
P=-\frac{\partial F^{(0)}}{\partial V}\equiv g^{(0)}(\beta,V).
\end{equation}
Then \re{5.14} gives a weakly-relativistic equation of state in the form
\begin{equation}\label{5.16}
P=g^{(0)}(\beta,V)-\frac3{\beta mc^2}\left((A+B)g^{(0)}+
A\beta\frac{\partial g^{(0)}}{\partial \beta}+
BV\frac {\partial g^{(0)}}{\partial V}\right),
\end{equation}
which is determined by the nonrelativistic form \re{5.15} and the
two constants, $A$ and $B$.

As an example we can consider a well known Van der Waals equation
\begin{equation}\label{5.17}
P=\frac N{\beta (V-Nb)}-\frac{N^2a}{V^2}
\end{equation}
with a pair of phenomenological constants, $a$ and $b$.
The corresponding weakly-relativistic equation is
\begin{equation}\label{5.18}
P=\frac N{\beta (V-Nb)}\left[1+\frac{3NbB}{\beta mc^2(V-Nb)}\right]
-\frac{N^2a}{V^2}\left[1+\frac{3(B-A)}{\beta mc^2}\right ].
\end{equation}
Up to the first order in $c^{-2}$ it can be presented in the nonrelativistic
form \re{5.17} with constants $a$ and $b$ being replaced by the
linear functions on temperature:
\begin{equation}\label{5.19}
a\mapsto a'=a\left(1+3\frac{B-A}{\beta mc^2}\right);\qquad
b\mapsto b'=b\left(1+\frac{3B}{\beta mc^2}\right).
\end{equation}
Corrections to other thermodynamical functions and a wider discussion
can be found in Ref.~\cite{[12]}.

\end{document}